\documentclass[12pt]{article}
\usepackage{hyperref}
\usepackage{graphicx}
\usepackage{xcolor}
\usepackage{dcolumn}
\usepackage{bm}
\usepackage{caption}
\usepackage{abstract}
\usepackage{amsmath,amssymb}
\title{\large\bf Geodesically Complete Regularized Schwarzschild Black Holes}

\author{Nihar Ranjan Ghosh\footnote{\tt g.nihar@iitg.ac.in}~ and
Malay K. Nandy\footnote{\tt mknandy@iitg.ac.in}\\
 {\em Department of Physics, Indian Institute of Technology Guwahati}\\ 
 Guwahati 781 039, India.
}

\date{April 20, 2025}

\begin{document}

\maketitle

\begin{abstract}
Classical general relativity predicts a singularity at the center of a black hole, where known laws of physics break down. This suggests the existence of deeper, yet unknown principles of Nature.  Among various theoretical possibilities, one of the most promising proposals is a transition to a de Sitter phase at high curvatures near the black hole center. This transition, originally proposed by Gliner and Sakharov, ensures the regularity of metric coefficients and avoids the singularity.  In search for such a regular black hole solution with finite curvature scalar, we propose a metric function $g_{rr}$ that exhibits a de Sitter-like core in the central region. An appealing feature of this metric is the existence of a {\em single} event horizon resembling the Schwarzschild black hole. Furthermore, the entire spacetime geometry is determined by the black hole mass alone, in agreement with the Isarel-Carter {\em no-hair theorem} for a charge-less, non-rotating black hole.  To determine the gravitational action consistent with such a solution, we consider a general Lagrangian density $f(\mathcal{R})$ in place of the Einstein-Hilbert action. By numerically solving the resulting field equation, we find that, in addition to the Einstein-Hilbert term, a Pad\'e approximant in the Ricci scalar $\mathcal{R}$ can produce such regular black hole solutions. To assess the physical viability of these black hole solutions, we verify that the proposed metric satisfies the principal energy conditions, namely, the dominant energy condition, weak energy condition, and  null energy condition, throughout the entire spacetime. Furthermore, in agreement with Zaslavskii's regularity criterion, the metric satisfies the strong energy condition in the range $r\geq r_h/2$, where $r_h$ is the event horizon. Furthermore, with the proposed regularized metric, the expansion scalar in the Raychaudhuri equation remains finite and its derivative vanishes at $r=0$, thereby preventing formation of caustic. This confirms that the spacetime is geodesically complete and free from true physical singularities.

\end{abstract}


\tableofcontents

\section{\label{Introduction}Introduction}

Black holes (BHs) are among the many important and beautiful outcomes of the general theory of relativity. It also serves as a theoretical laboratory to inspect the fundamental behaviour of the Nature. The first BH solution, the Schwarzschild solution \cite{schwarzschild1916gravitationsfeld}, came as a unique static vacuum solution to Einstein's field equations, $G_{ab}=8\pi  T_{ab}=0$. This solution gives the $4D$ metric $ds^2=-F(r)dt^2+F^{-1}(r)dr^2+r^2d\Omega^2$, with $F(r)=(1-2m/r)$, for a black hole of ADM mass $m$, and $d\Omega^2$ is the line element on a 2-sphere of unit radius. Although this metric is the first solution of the Einstein field equations describing the spacetime due to a massive gravitating body, it also gives rise to the fundamental question of what happens near the central singularity, as the Kretschmann scalar $R_{abcd}R^{abcd}$ diverges in an unbounded manner like $r^{-6}$ as $r\to 0$. 

 The divergence of the curvature scalar and the consequent breakdown of general relativity indicates some more rudimentary properties of a BH. One simple proposition is the transition, at some scale, of the spacetime itself to a de Sitter type geometry, with the hope that this core might help in regulating the divergence of the curvature scalar by giving some finite upper bound. 

This possibility required replacing the metric coefficient $g_{rr}$ of a BH with a de Sitter core at the centre with an eqaution of state $\rho=-p$, as first suggested  by Gliner and Sakharov \cite{gliner1966algebraic,sakharov1966initial}. Black holes with unique properties, such as, non-divergent metric coefficients and regular Kretschmann scalar as $r\to 0$, are generally known as regular black holes.
 
In this context, Wenda and Zhu \cite{wenda1988junction} proposed that the exterior geometry of a Schwarzschild BH can be joined with a singularity-free  de Sitter interior at the event horizon. But this direct matching of a de Sitter core with a exterior vacuum can never be carried out exactly at the horizon, as this will violate the continuity of pressure condition \cite{synge1960relativity}, $T^{ab}N_b=0 $, with $N_b$ the normal to the boundary.
However, although matching is not allowed at the horizon, but in principle this is allowed elsewhere \cite{poisson1988structure}.

Since the known laws of fundamental physics break down near the central singularity, 
it gives a possible indication that quantum effects might start dominating in the region near $r\approx0$ resulting in a singularity-free spacetime geometry. Quantum fluctuations might induce non-linear curvature $\mathcal{R}$ terms in the effective gravitational Lagrangian, that might remove the singularity. In addition to having a de Sitter type core in the centre, we hypothesize that modification of the Einstein-Hilbert action by replacing the Ricci scalar $\mathcal{R}$ with a functional $f(\mathcal{R})$ can result in a regular BH. In this study we adopt this scenario in search for a regularized black hole solution in $f(\mathcal{R})$ gravity.

To evaluate the physical plausibility and consistency of the proposed black hole solutions, we examine whether the associated spacetime metric satisfies the fundamental energy conditions commonly employed in general relativity. Specifically, we verify that the metric satisfies the dominant energy condition (DEC), the weak energy condition (WEC), and the null energy condition (NEC) throughout the entire spacetime manifold. These conditions ensure, respectively, that the energy density as measured by an observer is non-negative, that energy flows along causal trajectories, and that the stress-energy tensor behaves in a physically reasonable manner when contracted with null vectors.

In addition, we also consider the strong energy condition (SEC), which plays a central role in gravitational focusing and singularity theorems. While the SEC is often violated in regular black hole models due to the necessity of avoiding the singularity, we find that the proposed metric satisfies the SEC in the region $r \geq r_h/2 $, where $ r_h $ is the event horizon radius. This is consistent with the regularity criterion put forward by Zaslavskii \cite{zaslavskii2010regular}, which permits SEC violation near the core while maintaining regular and physically admissible behavior in the outer region. This compliance with the SEC provides further support to the physical viability of the regularized black hole metric.

Furthermore, within the framework of the proposed regularized metric, we confirm that the expansion scalar in the Raychaudhuri equation \cite{Raychaudhuri1955}, which typically governs the behavior of geodesic congruences, remains finite as $r$ approaches zero. In addition, the derivative of the expansion scalar vanishes at $r=0$, preventing the formation of caustics, which are usually associated with singularities in spacetime. This behavior is crucial because caustics correspond to regions where geodesics focus, leading to breakdown of physical laws. The absence of caustic formation under the regularized metric indicates that the spacetime is free from such singularities. This confirms that the spacetime is geodesically complete, meaning that all geodesics can be extended smoothly without encountering any singularities. This implies that no true BH singularity exists in this scenario.

The rest of the paper is organized as follows. In Section \ref{back}, a general background of regular BHs is given, that also includes the required regularity conditions on the black hole metric. Section \ref{Model} gives the proposed model for regular BHs and obtains its equivalent $f(\mathcal{R})$ gravity model. Furthermore, the energy conditions of the present model are discussed in Section \ref{NRG}. The Raychaudhuri equation is analyzed in Section \ref{rayc}. Finally, Section \ref{Discussion}  concludes the paper with a discussion.

\section{Background}
\label{back}

\subsection{Regular Black Holes}
The first regular BH solution without any central singularity was proposed by Bardeen \cite{bardeen1968proceedings}, with redshift function 
\begin{equation}
    \label{Bardeen}
    F(r)=1-\frac{2mr^2}{(r^2+q^2)^{3/2}}~,
\end{equation}
where $q$ is a magnetic charge and $m$ the ADM mass of the BH. 

Another regular BH solution of high importance was introduced by Hayward \cite{hayward2006formation}  by considering the Einstein tensor to be proportional to the square of the curvature.  A detailed discussion on regular BH can be found in \cite{de2010f}.

Bardeen's regular solutions can be found as a solution of Einstein gravity coupled with nonlinear electrodynamics (NLED) in the presence of an electromagnetic charge $q$ \cite{ayon2000bardeen, rodrigues2018bardeen}, where the central singularity is avoided by the self gravitating magnetic field.  

A general procedure of constructing a regular BH solution, in the presence of a magnetic or electric charge, by considering a coupling of the gravitational field with NLED has been offered in \cite{bronnikov2001regular, fan2016construction,bronnikov2017comment}. In these analyses, a two-parameter family of BH solutions were obtained, and by choosing a particular region of the parameter space, the singularity at the center was removed. This method for obtaining regular BHs can be generalized to include the cosmological constant and asymptotic AdS geometry. 

The above formalism can be extended further in the context of modified theories of gravity and NLED. Without specifying any definitive form for $f(\mathcal{R}) $ and the NLED Lagrangian, regular solutions were obtained by choosing an appropriate form for the mass function $m(r)$, as was shown in \cite{rodrigues2016generalisation}. Furthermore, such BHs have two horizons, namely an event horizon and a Cauchy horizon. Interestingly all energy conditions are satisfied throughout the spacetime, except for the strong energy condition, which is violated near the Cauchy horizon. Fortunately, it has been shown recently that, in general, the spacetime structure of a static, spherically symmetrical, regular BH violates the strong energy condition in any region inside the event horizon, in such a way that the Tolman mass becomes negative \cite{zaslavskii2010regular}.

Some of the other static regular BH solutions with electric and magnetic charge can be found in \cite{ayon1998regular,ayon1999new,ayon2005four,bronnikov2000comment,dymnikova2004regular,ansoldi2008spherical,balart2014regular,baalart2014regular,uchikata2012new,ponce2017regular,rodrigues2018using}, and with rotation in \cite{bambi2013rotating,neves2014regular,toshmatov2014rotating,azreg2014generating,dymnikova2015regular,torres2017regular} and some on alternative theories of gravity in \cite{berej2006regular,junior2015regular,rodrigues2016regular,rodrigues2019regular,de2018regular,junior2020regular,cano2021resolution}.

\subsection{Conditions on the redshift function}
\label{conditions}

We shall consider a static, spherically symmetric black hole spacetime with the metric given by 
\begin{equation}
    \label{general metric}
    ds^2=-F(r)dt^2+F^{-1}(r)dr^2+r^2(d\theta^2+\sin^2\theta d\phi^2)~,
\end{equation}
where $F(r)$ is designated as the redshift function.  
A surface of area $4\pi r^2$ is trapped if $F(r)<0$ and untrapped if $F(r)>0$, and the horizon of the BH is defined as the root of $F(r=r_h)=0$. The trapping horizons, in this case also the killing horizons, are located at $r=r_h$. 

In order that the spacetime is asymptotically Minkowskian, we require 
\begin{equation}
    \label{minkwski}
    F(r)\to 1~~~\text{as}~~~ r\to \infty~.
\end{equation}
Since we are interested in regularising a Schwarzschild BH, we further require a Schwarzschild-like spacetime for $r\gg r_h$, that is 
\begin{equation}
\label{schwarzschild}
    F(r)\sim 1-\frac{2m}{r} ~~~\text{for}~~~r\gg r_h,
\end{equation}
where $m$ is the ADM mass of the BH.

Furthermore, following Gliner, Sakharov, Bardeen, and Hayward \cite{gliner1966algebraic, sakharov1966initial, bardeen1968proceedings, hayward2006formation}, we regularize the BH interior with a de Sitter core,  requiring 
\begin{equation}
\label{cetral de sitter}
    F(r)\to 1-\frac{1}{3}\Lambda r^2 ~~\text{as} ~~r\to 0~,
\end{equation}
giving a non-divergent curvature at the centre of the BH.

A red shift function $F(r)$, satisfying the conditions \ref{minkwski}, \ref{schwarzschild}, and \ref{cetral de sitter}, is expected to give a regular Schwarzschild-like BH spacetime with ADM mass $m$.

\begin{figure}
    \centering
    \includegraphics[width=0.6\linewidth]{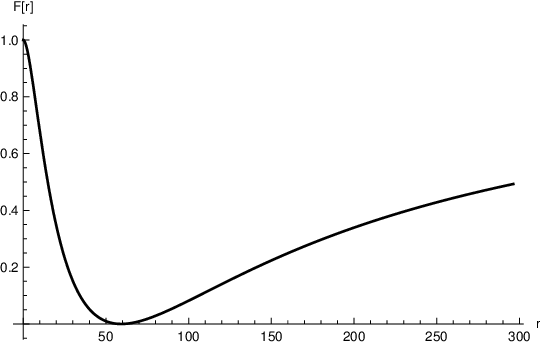}
    \caption{The red shift function $F(r)$ given by equation \ref{F(r)} for $m=100$, showing that $F(r)=0$ at the horizon $r=2 l=\frac{16}{27}m=59.2593$ of the regularised Schwarzschild BH. All numerical values are in Planck units, here and henceforth. }
    \label{F_vs_r}
\end{figure}

\section{Model for a regular Schwarzschild black hole}
\label{Model}

\subsection{{\label{F(r) model}}Regularisation of the singularity}

We regularize the BH spacetime by considering a simple model for the redshift function $F(r)$, written as
\begin{equation}
    \label{F(r)}
    F(r)=1-\frac{2mr^2}{(r+l)^3}~.
\end{equation}
Clearly, this function satisfies all three conditions \ref{minkwski}, \ref{schwarzschild}, and \ref{cetral de sitter}, in particular
\begin{equation}
    \begin{split}
        F(r)&\sim1-\frac{2m}{r}~~~\text{for}~~~ r\gg r_h,\\
        &\sim 1-\frac{1}{3}\Lambda r^2~~~ \text{for}~~~r\ll l,
    \end{split}
\end{equation}
where $\Lambda=6m/l^3$.

The horizons are given by $F(r)=0$, giving three roots. It turns out that physically acceptable solutions exist if and only if 
\begin{equation}
\label{lm}
l=\frac{8}{27}m~,
\end{equation}
in which case one root is negative ($r_1=-\frac{1}{4}l$) and the other two roots are equal and positive ($r_2=r_3=+2l$), thus defining a {\em single} event horizon at radius
\begin{equation}
r_h=2l~.
\end{equation}
This property of the metric  \ref{F(r)} having a {\em single} event horizon coincides with the nature of spacetime of {\em Schwarzschild black hole}. 

It is important to note that a slight departure from relation \ref{lm} would lead to unphysical solutions of $F(r)=0$. For example, with $l=\frac{8.1}{27}m$, the roots are  $-0.075 m$,  $(0.588 - 0.115 i) m$,  $(0.588 + 0.115 i) m$. Similarly, for $l=\frac{7.9}{27}m$, the roots are complex with very small imaginary parts. Thus, $l=\frac{8}{27}m$ is the only relation that can give a physical horizon.

The metric parameter $l$ being related with the mass $m$ via \ref{lm}, the entire spacetime geometry is determined by the mass of the black hole alone. This is in agreement with the Isarel-Carter {\em no-hair theorem} \cite{Israel1967, Israel1968, Carter1971} for a charge-less, non-rotating black hole, such as the Schwarzschild black hole.

Figure \ref{F_vs_r} displays the profile of the redshift function $F(r)$, where the zero of the function defines the black hole horizon.

\begin{figure}
    \centering
    \includegraphics[width=0.6\linewidth]{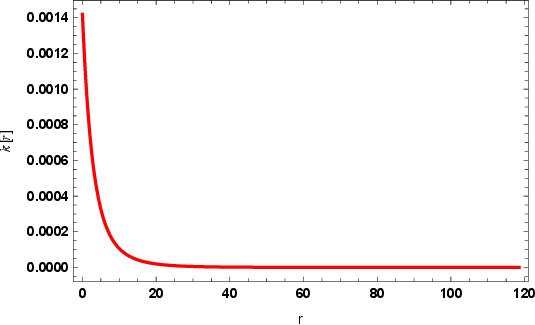}
    \caption{Radial profile of the Kretschmann scalar $\mathcal{K}(r) $ given by equation \ref{kretschmann} for $m=100$. At the origin $r=0$, the Kretschmann scalar has a finite value $\frac{96m^2}{l^6}=96\times(\frac{27}{8})^6\frac{1}{m^4}=0.001419$, indicating a regular spacetime inside $(r<2l=59.2593)$ of the regularised Schwarzschild BH. }
    \label{kretschfig}
\end{figure}

The metric \ref{general metric}, with the red shift function \ref{F(r)}, yields the 
Ricci scalar $\mathcal{R}$ as
\begin{equation}
    \label{Ricci scalar}
    \mathcal{R}=\frac{24l^2m}{(r+l)^5}
\end{equation} 
\begin{figure}
    \centering
    \includegraphics[width=0.7\linewidth]{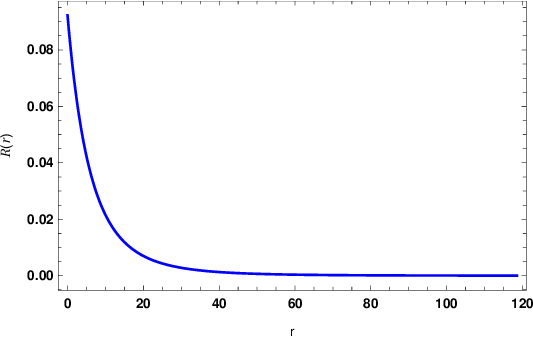}
    \caption{Radial profile of the Ricci scalar $\mathcal{R}(r) $ given by equation \ref{Ricci scalar} for $m=100$. At the origin $r=0$, the Ricci scalar has a finite value $\frac{24m}{l^3}=24\times(\frac{27}{8})^3\frac{1}{m^2}=0.09226$, indicating a regular spacetime inside $(r<2l=59.2593)$ of the regularised Schwarzschild BH. }
    \label{Ricci}
\end{figure} 
and the Kretschmann scalar
\begin{equation}
    \label{kretschmann}
    \mathcal{K}=R_{abcd}R^{abcd}=48m^2\frac{(2l^4+7l^2r^2-2lr^3+r^4)}{(r+l)^{10}}~.
\end{equation}
Thus the Ricci scalar $\mathcal{R}$ as well as the Kretschmann scalar $\mathcal{K}$ are finite and regular upon approaching the center, $r\to0$. These features indicate regularity of the BH spacetime at the center. Moreover, we have $\mathcal{R}\to0$  and $\mathcal{K}\to0$ as $r\to\infty$,
 showing asymptomatic flatness of the spacetime.  
 Figures \ref{Ricci} and \ref{kretschfig}  illustrate these features clearly.

The metric \ref{general metric}, with the red shift function \ref{F(r)}, yields the components of the Einstein tensor $G_{ab}=R_{ab}-\frac{1}{2}g_{ab}R$, as
\begin{equation}
    \label{einstein tensor components}
    \begin{split}
        G_t^t&=G_r^r=-\frac{6lm}{(r+l)^4}~,\\
        G_\theta^\theta&=G_\phi^\phi=\frac{6lm(r-l)}{(r+l)^5},
    \end{split}
\end{equation}
showing that all the coefficients falls off very rapidly, $\mathcal{O}(r^{-4})$ for $r\gg l$.

These components of the Einstein tensor can be used to relate with the components of the effective energy-momentum tensor $\Theta_{ab}$, defined by $G_{ab}=\Theta_{ab}$.  Thus, the energy density $-\Theta_t^t$ is given by
\begin{equation}
    \label{density}
    -G_t^t=-\Theta_t^t=\frac{6l}{m^{1/3}}\left(\frac{E(r)^\frac{1}{3}}{r}\right)^4
\end{equation}
and
\begin{equation}
    \label{energy}
    g^{rr}=1-\frac{2E(r)}{r}~,
\end{equation}
where the energy $E(r)$ is defined by
\begin{equation}
    \label{energy1}
    E(r)=\frac{m}{(1+l/r)^3}~.
\end{equation}

Equations \ref{Ricci scalar}, \ref{density} and \ref{energy1} suggest that $G^t_t\sim \mathcal{R}^{4/5}$. Thus assuming $G^{ab} =\Theta^{ab}\sim \mathcal{R}^{4/5}$, it can be shown that the component $dE/dr=-(1/2)r^2\Theta^t_t$ of the Einstein equations yields a simple singularity-free solution for the metric coefficient $g^{rr}$ in the form \ref{F(r)} with appropriate boundary conditions.

\subsection{{\label{f(R) model}}The equivalent $f(\mathcal{R})$ gravity model}

In this work, we shall be interested in constructing an equivalent $f(\mathcal{R})$ theory of gravity in the Jordan frame that mimics the above regularized spacetime geometry.

Consequently, we begin by considering the general action 
\begin{equation}
    \label{action}
    \mathcal{A}=\frac{1}{2}M_p^2\int d^4x\sqrt{-g}f(\mathcal{R}),
\end{equation}
where $g=\text{det}[g_{ab}]$, $g_{ab}$ being the metric of the spacetime, given by equation \ref{general metric}. We shall take the regularised red shift function $F(r)$ defined in equation \ref{F(r)}.

Extremum of the action \ref{action} yields the modified field equation as
\begin{equation}
    \label{f(R) dynamicl equn 1}
    f'(\mathcal{R})R_{ab}-\frac{1}{2}f(\mathcal{R})g_{ab}=\left[\nabla_a\nabla_b-g_{ab}\square  \right]f'(\mathcal{R}),
\end{equation}
where $f'(\mathcal{R})=\frac{df}{d\mathcal{R}}$.

The Ricci tensor $R_{ab}$ can be obtained from equation \ref{f(R) dynamicl equn 1} in the form
\begin{equation}
\label{fR1}
R_{ab}= \frac{1}{f'(\mathcal{R})}\Big[\frac{1}{2}f(\mathcal{R})g_{ab}+\left[\nabla_a\nabla_b-g_{ab}\square  \right]f'(\mathcal{R})\Big] ~, 
\end{equation}
which is the Ricci scalar in vacuum in the $f(\mathcal{R})$ model.

On the other hand, when we regularised the BH metric in the previous section, we obtained the field equation as $R_{ab}-\frac{1}{2}\mathcal{R}g_{ab}=\Theta_{ab}$, or equivalently  
\begin{equation}
\label{fR2}
 R_{ab}=\Theta_{ab}-\frac{1}{2}\Theta g_{ab} ~. 
\end{equation}
Equation \ref{fR1} represents the Ricci tensor of the vacuum spacetime in $f(\mathcal{R})$ gravity. Consequently, in order to find an equivalent $f(\mathcal{R})$ representation for the regularised BH, we obtain from equation \ref{fR2},   
\begin{equation}
\label{fR3}
\frac{1}{f'(\mathcal{R})}\Big[\frac{1}{2}f(\mathcal{R})g_{ab}+\left[\nabla_a\nabla_b-g_{ab}\square  \right]f'(\mathcal{R})\Big] =\Theta_{ab}-\frac{1}{2}\Theta g_{ab} ~.  
\end{equation}
Taking trace on both sides, and using the fact that $\mathcal{R}=-\Theta$, we have
\begin{equation}
    \label{box f}
    \square f'(\mathcal{R})=\frac{1}{3}\Big[2f(\mathcal{R})-\mathcal{R} f'(\mathcal{R})\Big]~.
\end{equation}
Substitution of equation \ref{box f} in equation \ref{f(R) dynamicl equn 1} gives
\begin{equation}
    \label{finl dynamicl eqn}
    f'(\mathcal{R})R_{ab}=\frac{1}{3}g_{ab}\Big[\mathcal{R} f'(\mathcal{R})-\frac{1}{2}f(\mathcal{R})\Big]+\nabla_a\nabla_b f'(\mathcal{R}),
\end{equation}

\begin{figure}
    \centering
    \includegraphics[width=0.7\linewidth]{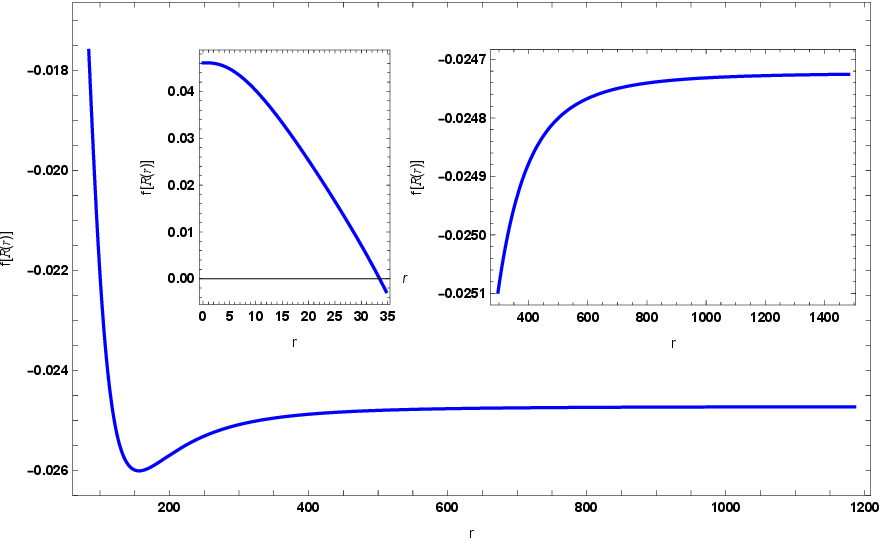}
    \caption{Radial profile of the Lagrangian $f[\mathcal{R}(r)]$ in the modified gravity action \ref{action}, obtained by numerical integration of the differential equation \ref{numerical eqn} with initial conditions $f[\mathcal{R}(r=0)]=\frac{12m}{l^3}=12\times(\frac{27}{8})^3\frac{1}{m^2}=0.046132$ for $m=100$ and $f'[\mathcal{R}(r=0)]=0$. The first inset shows its behaviour for small radial coordinates where it crosses over to negative values at $r\approx33.5$, which is slightly higher than $l=\frac{8}{27}m=29.6296$. The second inset shows that it approaches zero for large radial coordinates.}
    \label{f(R)_vsr}
\end{figure}

Equation \ref{Ricci scalar} expresses the Ricci scalar as a function of the radial coordinate, $\mathcal{R}(r)$. We can thus express the derivative as
\begin{equation}
f'[\mathcal{R}(r)]=\frac{df}{d\mathcal{R}}=-\frac{(r+l)^6}{120l^2m}~\frac{df}{dr}~.
 \end{equation}
Consequently,  the ``$\theta\theta$'' or ``$\phi\phi$'' component of equation \ref{finl dynamicl eqn} leads to
\begin{equation}
    \label{numerical eqn}
    A(r)\frac{d^2f}{dr^2}+B(r)\frac{df}{dr}+C(r)f=0,
\end{equation}
with
\begin{equation}
\label{coefficients}
    \begin{split}
        A(r)&=(r+l)^6F(r)~,\\
        B(r)&=2lmr(r+l)(3r-l)+6(r+l)^5F(r)~,\\
        C(r)&=-20l^2mr~,
    \end{split}
\end{equation}
where $F(r)$ is the regularised red shift function given by equation \ref{F(r)}.

\subsection{Numerical Integration}

We numerically integrate the differential equation \ref{numerical eqn} with initial conditions $f[\mathcal{R}(r=0)]=\frac{12m}{l^3}$ and $f'[\mathcal{R}(r=0)]=0$ as appropriate for our model in order to obtain $f[\mathcal{R}(r)]$ as a function of the radial coordinate $r$ in the range $r\in[0,\infty)$.

Figure \ref{f(R)_vsr} shows the radial profile of $f[\mathcal{R}(r)]$. It is clear from the plot that in the vicinity of the origin, $r\approx0$, the behaviour of $f(\mathcal{R})$ differs significantly from Einstein gravity $f(\mathcal{R})=\mathcal{R}$, and it asymptotically approaches zero at long distances. Therefore, our choice of the redshift function $F(r)$ given by \ref{F(r)} modifies the internal structure of the BH appropriately, by replacing the central singularity at  $r=0$ with a de Sitter core while keeping the long distance $(r\gg 2l)$ behaviour the same as that of a Schwarzschild BH with ADM mass $m$.

To find the dependency of $f[\mathcal{R}]$, shown in figure \ref{f(R)_vsr}, as a function of the Ricci scalar $\mathcal{R}$, we obtain the best fit by using the Pad\'e approximant, given by
\begin{equation}
    \label{pade}
    f[\mathcal{R}]=a_0\mathcal{R}+ \frac{a_0-a_1 r+a_2 r^2+a_3 r^3-a_4r^4}{1-b_1 r+b_2 r^2-b_3 r^3+b_4r^4}~,
\end{equation}
where the radial coordinate is expressed as a function of $\mathcal{R}$ by using equation \ref{Ricci scalar} as
\begin{equation}
    \label{pade1}
    r=\left(\frac{24l^2m}{\mathcal{R}}\right)^{1/5}-l~,
\end{equation}
with $a_0\gg a_1\gg\dots\gg a_4 $, and $b_1\gg b_2\gg\dots\gg b_4 $, and $a_i,b_i>0$. The first term on the RHS in equation \ref{pade} implies the Einstein-Hilbert Lagrangian density with a constant prefactor.

\section{Energy Conditions}
\label{NRG}

We further investigate if the metric \ref{general metric} with the redshift function \ref{F(r)} satisfies the four energy conditions: the Weak Energy Condition (WEC), Null Energy Condition (NEC), Strong Energy Condition (SEC) and Dominant Energy Condition (DEC) \cite{padmanabhan2010gravitation, Poisson_2004, carroll2019spacetime}.

Consider a timelike four velocity $t^\mu$ and null-like four velocity $n^\mu$. For a stationary observer in the comoving frame $t^\mu=[1/\sqrt{F},0,0,0]$ and for a null radial trajectory $n^\mu=[1,F,0,0]$, satisfying $t^\mu t_\mu=-1$ and $n^\mu n_\mu=0$. Then using the methodology shown in \cite{santos2007energy}, we have for SEC
\begin{equation}
    \label{SEC}
    R_{\mu\nu}t^\mu t^\nu=\frac{6lm(r-l)}{(r+l)^5}~\geq 0,~~~~\forall~~r\in [l,\infty)
\end{equation}
which through the Raychaudhuri equation suggests that the proposed gravity model is attractive. Clearly the SEC is satisfied only in the range $r\in [l,\infty)$. However, the SEC is violated in the range $r\in[0,l]$, which is consistent with the findings of \cite{zaslavskii2010regular}, which states that for any static, spherically symmetric regular BH, the SEC is violated for any region inside the event horizon.   

\begin{figure}
    \centering
    \includegraphics[width=0.6\linewidth]{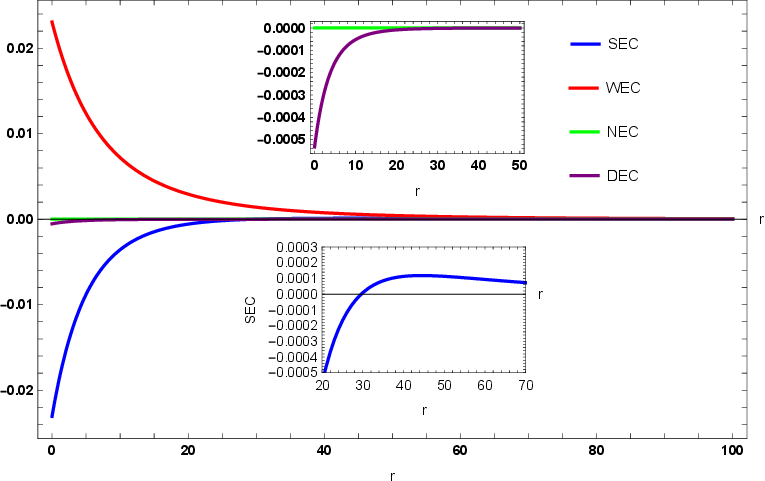}
    \caption{Radial profiles of SEC, WEC, DEC, NEC given by equations \ref{SEC}, \ref{WEC}, \ref{DEC}, \ref{NEC} with the regularised red shift function \ref{F(r)} for $m=100$.}
    \label{enrgy conditn}
\end{figure}

On the other hand, the WEC is satisfied in the entire region of the spacetime,
\begin{equation}
    \label{WEC}
    \left[R_{\mu\nu}-\frac{1}{2}Rg_{\mu\nu}  \right]t^\mu t^\nu=\frac{6lm}{(r+l)^4}~\geq 0,~~~~\forall~~r\in [0,\infty)~,
\end{equation}
indicating that the energy density measured by any observer is always non-negative, in agreement with Hawking and Ellis \cite{hawking2023large}.

The DEC is also satisfied in the entire region of the spacetime,defining a single event
\begin{equation}
    \label{DEC}
    \begin{split}
        \Big[R_{\mu\nu}-&\frac{1}{2}Rg_{\mu\nu}  \Big]\Big[R_{\lambda\rho}-\frac{1}{2}Rg_{\lambda\rho}  \Big]g^{\nu\rho}t^\mu t^\lambda\\
        &=-\left[\frac{6lm}{(r+l)^4}\right]
        \leq 0~,~~~~\forall~r\in [0,\infty)~,
    \end{split}
\end{equation}
which implies non-negative energy density in addition of having a local energy flow vector which is non-spacelike. 

Furthermore, the NEC is satisfied throughout the spacetime,
\begin{equation}
    \label{NEC}
    \Big[R_{\mu\nu}-\frac{1}{2}Rg_{\mu\nu}  \Big]n^\mu n^\nu=0~,~\forall~r\in [0,\infty)~,
\end{equation}
having its fundamental importance in signifying boundedness from below of the corresponding Hamiltonian.

Figure \ref{enrgy conditn} illustrates radial profiles of the above energy conditions \ref{SEC}, \ref{WEC}, \ref{DEC}, and \ref{NEC} derived from the red shift function $F(r)$ given by equation \ref{F(r)}. The figure clearly shows that WEC, NEC and  DEC are satisfied throughout the spacetime, $r\in [0,\infty)$. On the other hand, SEC holds only in the range $r\in [l,\infty)$, whereas its violation in the range $r\in [0,l]$ is acceptable as it occurs inside the event horizon \cite{zaslavskii2010regular}.

\section{Raychaudhuri Equation}
\label{rayc}

To conclude the work, we verify if the singularity at $r=0$ is actually resolved by looking for the formation of caustics in the framework of the Raychaudhuri equation \cite{Raychaudhuri1955}. The standard timelike Raychaudhuri equation, for an affinely parametrized hypersurface orthogonal $(\omega_{ab}\omega^{ab}=0)$ congruences is 
\begin{equation}
    \label{Ray}
    \frac{d\theta}{d\tau}=-\frac{1}{3}\theta^2-2\sigma^2-R_{ab}u^au^b~,
\end{equation}
with $\theta$ the expansion scalar, $\tau$ the proper time, and $2\sigma^2=\sigma^{ab}\sigma_{ab}$ the shear tensor associated with the congruence defined by the timelike vector field $u^a$.

To analyze this equation,  we consider a radial and marginally bounded timelike congruence, that gives the energy $E$ as
\begin{equation}
    E=-u_a\zeta^a=-u_t=1~,
\end{equation}
for $\zeta^a_{(t)}=(1,0,0,0)$, with $u^a$ the four velocity of the timelike geodesic. With our metric, this turns out to be
\begin{equation}
    u^a=\left[\frac{1}{F(r)}, \epsilon\sqrt{\frac{2m r^2}{(r+l)^3}},0,0  \right]
\end{equation}
with $\epsilon=+1~(-1)$ indicating the outgoing (ingoing) geodesics.

Thus, the expansion scalar $\theta$ is found to be
\begin{equation}
    \label{theta}
        \theta(r)=\nabla_au^a=\epsilon\sqrt{\frac{9m}{2}}\frac{(2l+r)}{(r+l)^{5/2}}~.
\end{equation}
It is obvious from equation \ref{theta} that at $r=0$ the expansion scalar $\theta$ is non-zero negative (positive) for an ingoing (outgoing) timelike congruence. 

The rate of change of the expansion scalar $d\theta/d\tau$ is thus readily obtained from equation \ref{theta} as
\begin{equation}
    \label{dtheta}
    \frac{d\theta}{d\tau}=-\frac{3m}{2}\frac{r(8l+3r)}{(r+l)^5}~.
\end{equation}

In the Schwarzschild spacetime, for a radial and marginally bounded congruence with four velocity $u^a=[(1-2m/r)^{-1},\epsilon\sqrt{2m/r},0,0]$, it can be shown that the expansion scalar is
\begin{equation}
    \label{sch_theta}
    \theta=\epsilon\frac{3}{2}\sqrt{\frac{2m}{r^3}}~,
\end{equation}
giving
\begin{equation}
    \label{sch_dtheta}
    \frac{d\theta}{d\tau}=-\frac{9m}{2r^3}~.
\end{equation}

\begin{figure*}
\includegraphics[width=0.45\textwidth]{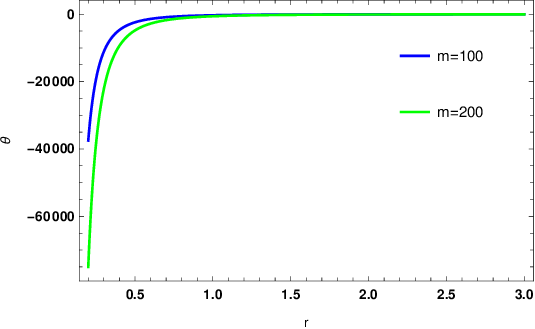}
\includegraphics[width=0.45\textwidth]{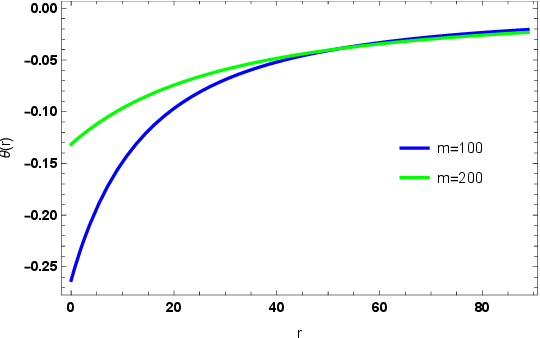}
\caption{Left Panel: Radial profile of the expansion scalar $\theta(r)$ for Schwarzschild BHs, for two different masses, $m=100,~200$. The expansion scalar approaches negative {\em infinity} as $r\to 0$, indicating infinite focusing and formation of caustic. Right Panel: Radial profile of the expansion scalar $\theta(r)$ for regularized Schwarzschild BHs, for two different masses, $m=100,~200$. The expansion scalar approaches negative {\em finite} values as $r\to 0$ for both masses, showing that no caustic is formed inside the BHs at $r=0$.}
\label{theta graph}
\end{figure*}

Thus, for the Schwarzschild case, both $\theta$ and $d\theta/d\tau$ diverge at $r=0$, indicating an infinite focusing and formation of caustic.
 
In contrast, with the present model of nonsingular metric, equations \ref{theta} and \ref{dtheta} indicate nondiverging expansion scalar and its rate at the center, namely, 
\begin{equation}
\theta(r=0)=\epsilon\sqrt{\frac{18m}{l^3}}~~\text{and}~~ \frac{d\theta}{d\tau}\Big|_{r=0}=0. 
\end{equation}
 
It is obvious that for $l=0$, just like the regularized metric lead to the  Schwarzschild geometry with singularity at the centre, equations \ref{theta} and \ref{dtheta} 
lead to equations \ref{sch_theta} and \ref{sch_dtheta} respectively, as expected. Thus, the parameter $l$, related to the mass as $l=\frac{8}{27}m$, acts as a regulator in the divergence of $\theta$ and $d\theta/d\tau$.

Figure \ref{theta graph} shows the behaviour of expansion scalar $\theta
$ as a function of coordinate $r$ for the Schwarzschild BH and for the regularized BH metric, respectively. The left panel shows that the expansion scalar diverges as one approaches $r=0$. On the other hand, for the regular metric considered here, it is clear from the right panel that the expansion scalar reaches a finite negative value upon approaching $r=0$.

\begin{figure*}
         \includegraphics[width=0.45\textwidth]{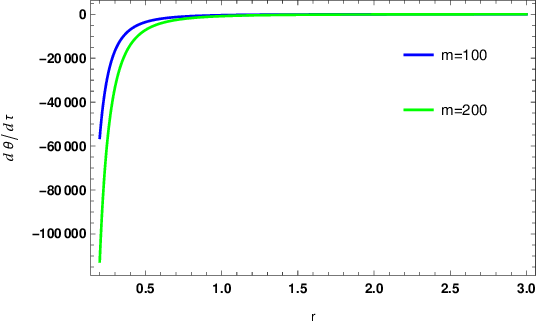}
         \includegraphics[width=0.45\textwidth]{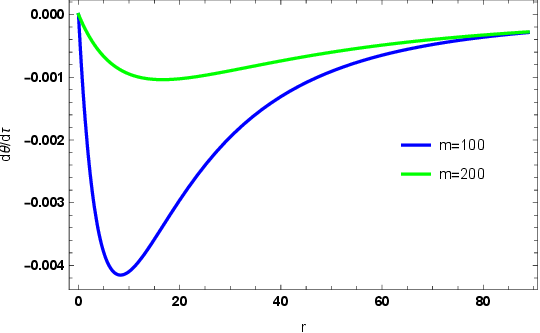}
         \caption{Left Panel: Radial profiles of $\frac{d\theta}{d\tau}$ for Schwarzschild BHs for two different masses $m=100, ~200$. In both cases, $\frac{d\theta}{d\tau}$ approaches negative {\em infinity} as $r\to 0$, indicating a singularity at $r=0$. Right Panel: Radial profiles of $\frac{d\theta}{d\tau}$ for the regularized Schwarzschild BHs for two masses $m=100, ~200$. In each case, the behavior of $\frac{d\theta}{d\tau}$ indicates a bounceback from a finite negative value at  $r=0.280464l$, showing a true singularity-free center inside the BH. }
        \label{dtheta graph}
\end{figure*}

Figure \ref{dtheta graph} shows the behaviour of $d\theta/d\tau$ for Schwarzschild BH and the regularized metric, as a function of the coordinate $r$. The left panel shows that $d\theta/d\tau$ diverges as one approaches $r=0$, and the strength of divergence increases with the increase in mass for Schwarzschild BH. On the other hand, the right panel shows that, apart from being always finite and negative indicating attractive gravitational field, $d\theta
/d\tau$ bounces back from a finite negative value
and goes to zero at $r=0$. The right panel further shows that the bouncing back happens much earlier for massive BHs, as was shown in \cite{PhysRevD.103.084038}, using modified Raychaudhuri equation in loop quantum gravity. Interestingly, the plot for $d\theta/d\tau$ for our regular metric, near $r\approx 0$ looks very similar to that obtained in \cite{PhysRevD.103.084038} from loop quantum gravity corrections to Raychaudhuri equation.

The minima of the $\frac{d\theta}{d\tau}$ versus $r$ plot corresponds to the point where the bouncing happens.  Differentiating equation \ref{dtheta} with respect to $r$ and equating it to zero, it can be shown that the bouncing happens at $r=0.280464l$, which is well inside the region $r\in(0,l)$ where the SEC is violated.
 
We see that the first two terms on the RHS of equation \ref{Ray} are always negative, therefore always contributing to the focusing of the geodesics. However, the quantity $R_{ab}u^au^b$ present in the last term changes its sign when $r<l$, and the rate of focusing starts to decrease as the congruence falls inside the SEC violating region. This last term becomes more and more dominant as one approaches $r=0$, which ultimately leads to a bounce. Therefore, the term $R_{ab}u^au^b$, which also violates the SEC when $r<l$, helps in the prevention of caustic formation at $r=0$.

We close this discussion with the following remarks.
The nature of singularities in general relativity is much different from those in other field theories. Although divergences of different curvature scalars indicate the possible existence of a singularity in the spacetime, this divergence alone is not sufficient. The necessary and sufficient condition for a singular spacetime is the existence of a congruence which begin and end at a finite affine parameter, the proper time, say. Such a spacetime is called geodesically incomplete. 
Furthermore, Hawking and Penrose have shown that all spacetime solutions of Einstein Gravity will definitely be singular \cite{PhysRevLett.14.57, hawking1970singularities}. This conclusion requires that there exists a congruence such that they focus to form caustic at a finite proper time. It is clear from equations \ref{theta} and \ref{dtheta} that  the expansion scalar is non-zero finite and $d\theta/d\tau=0$ at $r=0$, indicating that no caustic can formed at $r=0$. This confirms that the considered regularized Schwarzschild metric is geodesically complete.

\section{Discussion and Conclusion}
\label{Discussion}

In this work, we explored a regular black hole metric characterized by a non-diverging metric coefficients at \( r = 0 \), as described by Equation \ref{F(r)}. This specific metric structure exhibits a de Sitter-like core at small radial distances, with an effective cosmological constant given by \( \Lambda = 6m/l^3 \). At larger distances, the metric smoothly transitions into a Schwarzschild-like geometry, ensuring compatibility with well-established results in the weak-field limit.  

An attractive feature of this metric model is the existence of a {\em single} event horizon resembling the Schwarzschild black hole. Furthermore, the metric parameter $l$ is found to be related with the mass $m$ of the black hole, via \ref{lm}. This is in agreement with the Isarel-Carter {\em no-hair theorem} for a charge-less, non-rotating black hole.

One of the crucial aspects of constructing a physically meaningful regular black hole solution is ensuring that all curvature invariants remain finite throughout the spacetime. In our model, we verified that both the Ricci scalar \( \mathcal{R} \) and the Kretschmann scalar \( R_{abcd} R^{abcd} \) remain well-behaved and finite for all \( r \in [0, \infty) \). This confirms that the proposed metric successfully removes the central singularity, replacing it with a regular core consistent with fundamental principles of physics.  

To determine the underlying gravitational action that gives rise to such a regular black hole solution, we adopted a general \( f(\mathcal{R}) \) gravity framework, in place of the Einstein-Hilbert action. 
By demanding that the gravitational field equations arising from this action leads to our regular black hole metric in vacuum, we obtained a differential equation (\ref{numerical eqn}) that the function \( f(\mathcal{R}) \) must satisfy to generate such a regular black hole solution.  

A significant outcome of this analysis is the specific form of \( f(\mathcal{R}) \), expressed as a Pad\'e approximant in the Ricci scalar. Our numerical results, presented in Figure \ref{f(R)_vsr}, confirm that the function \( f(\mathcal{R}) \) starts from a de Sitter core at small \( r \) and smoothly transitions to an asymptotically flat spacetime. A best-fit analysis reveals that the leading-order term in \( f(\mathcal{R}) \) corresponds to the Einstein-Hilbert Lagrangian density \( \mathcal{R} \), consistent with the fact that general relativity remains valid asymptotically in the large distance limit.  

Apart from constructing a regular black hole solution, it is essential to verify the energy conditions to assess physical viability of such black holes. We therefore confirmed that the proposed regular metric satisfies the principal energy conditions.  Our numerical analysis, illustrated in Figure \ref{enrgy conditn}, shows that the weak energy condition (WEC), the null energy condition (NEC), the dominant energy condition (DEC) hold throughout the entire spacetime. The strong energy condition (SEC) is satisfied in the range \( r \geq r_h/2 \), which is consistent with Zaslavskii’s criterion, that confirms that SEC can be violated anywhere inside the event horizon for regular black holes \cite{zaslavskii2010regular}. These energy conditions further support the physical credibility of our regularized metric model.

In the context of singularity resolution, it is important to note that absence of divergence in curvature scalars may not be a true indicator of the absence of singularity in the spacetime. A spacetime is truly singular only if it is geodesically incomplete, meaning some timelike or null geodesics terminate at a finite affine parameter (proper time, say). Hawking and Penrose established that under general conditions, spacetime solutions in Einstein's gravity inevitably contain such singularities due to geodesic focusing. However, in the case of the regularized Schwarzschild metric examined here, the expansion scalar remains finite and its derivative vanishes at $r = 0$, precluding the formation of caustics. This demonstrates that the black hole spacetime is {\em geodesically complete}, and thus free from singularity at the center.

Overall, our findings demonstrate that it is possible to construct a physically viable regular black hole solution with a {\em single} event horizon resembling the Schwarzschild BH with a de Sitter core in the \( f[\mathcal{R}] \) gravity framework.
This approach not only eliminates the central singularity but also ensures consistency with energy conditions, as well as agreement with the Isarel-Carter {\em no-hair theorem}.

\subsection*{Acknowledgment}
Nihar Ranjan Ghosh is supported by  the Ministry of Human Resource Development, Government of India.


\end{document}